\documentclass[twocolumn,showpacs,showkeys]{revtex4}
\usepackage{graphicx}
\usepackage{bm}
\usepackage{color}
\usepackage{amsmath}
\usepackage{natbib}

\begin{document}

\title{Influence of finite size of molecules on spectrum of fully polarized dipolar BECs: Generalization of Bogoliubov's spectrum}

\author{Pavel A. Andreev}
\email{andreevpa@physics.msu.ru}
\author{L. S. Kuz'menkov}%
\email{lsk@phys.msu.ru}
\affiliation{M. V. Lomonosov Moscow State University, Moscow, Russian Federation.}

 \date{\today}

\begin{abstract}
We discuss complete theory of point-like particles in fully polarized BECs describing difference in behaviour of electrically- and magnetically polarized BECs. Next we present generalization of this theory on finite size particles to include contribution of size of real molecules on dynamical properties of dipolar BECs. As an application of obtained equation we calculate spectrum of linear collective excitations getting generalization of the Bogoliubov's spectrum.
We show absence of roton instability due to positivity of polarization contribution in the excitation spectrum. We found that new type of instabilities appears in small wave length limit due to finite size of molecules.
 \end{abstract}


\maketitle


Theoretical research of dipolar Bose-Einstein condensates (BECs) has been a bright topic for the last fourteen years. Many interesting properties of dipolar BECs related to the long range dipole-dipole interaction between particles give rise to experimental efforts for creation of dipolar BECs. Two kinds of dipolar particles exist. The first of them is atoms with large magnetic moment ($^{52}Cr$ is an example \cite{Lahaye Nature 07}). A lot of efforts have been applied for creation of electrically dipolar BECs of molecules, which are the second kind of dipolar BECs. The electrically polarized BECs have not been made yet, but their creation has been expected, since BECs of molecules should reveal strong dipolar properties \cite{Giamarchi NP 08}, \cite{Ni PCCP 09}, \cite{Carr NJP 09}. A generalization of the Gross-Pitaevskii equation has been used for theoretical research of the dipolar BECs \cite{Yi PRA 00}-\cite{Yi PRA 01}. The generalized Gross-Pitaevskii (gGP) equation contains the long-range dipole-dipole interaction along with the short-range interaction existing in the Gross-Pitaevskii equation.

Steady interest exists in the field of dipolar BECs \cite{Jona-Lasinio 13}-\cite{Wilson PRA 12} that bring us to give proper examination of basic theory in the field. Hence correct mean-field theory of dipolar BECs is under discussion in this letter. The multi-disciplinary method of many-particle quantum hydrodynamics is used for microscopic justification of corresponding equations. Usually one considers point-like particles at studying of systems with the long-range interaction. However electric dipolar molecules have size of order of 200-500 pm. Thus account of the particle size can give considerable contribution in collective physical effects. Getting finite size of particles instead of the point-like particles we present generalized mean-field theory developing general approach that can be used in different fields of condensed matter and plasma physics.

\begin{figure}
\includegraphics[width=8cm,angle=0]{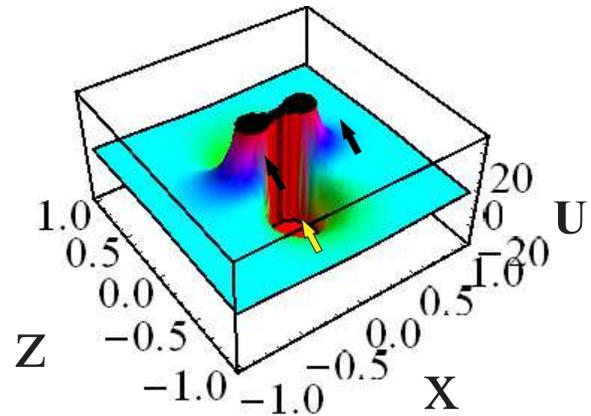}
\caption{\label{ESS 1} (Color online) The figure shows the
potential of dipole-dipole interaction for parallel dipoles illustrating attraction in direction parallel to dipoles and repulsion in direction perpendicular to dipoles. We draw a plane $y=0$. The arrow in the center of the picture presents one of the dipole considered as the source of the field. Two other arrows illustrates dipoles in two different positions. Dark (white) arrow presents a situation then $\textbf{r}-\textbf{r}'$ is perpendicular (parallel) $\textbf{d}$.}
\end{figure}

Since the gGP equation contains the potential of dipole-dipole interaction under the integral over whole space it is unwise to neglect delta-function term in full potential of dipole-dipole interaction. The delta-function term was restored in our recent papers \cite{Andreev 2013 non-int GP}, \cite{Andreev EPJ D Pol}, \cite{Andreev 2013 Dip+Spin}. We have shown that the account of delta-function terms for electrically- and magnetically polarized BECs reveals difference in dynamical properties of the two kinds of BECs. We have shown that electromagnetic field created by particles of dipolar BECs satisfies the Maxwell equations at consideration of the delta-function terms. Moreover we have different pairs of the Maxwell equations for the different kinds of BECs that reveals the fundamental difference of the two kinds of dipolar BECs. This difference takes place if we consider model of point-like particles. The question "How properties of dipolar BECs change at consideration of finite size of particles?" in the main topic of this letter.

Majority of papers include the potential energy of dipole-dipole interaction in the following form
\begin{equation}\label{FSDbec d-d Ham fraction
only} U_{dd}=\frac{\textbf{d}^{2}-3(\textbf{d}\textbf{r})^{2}/r^{2}}{r^{3}},\end{equation}
which is depicted on Fig. (1).
However we have to account the delta-function term for the point-like particles
\begin{equation}\label{FSDbec d-d Ham full} U_{dd_{full}}=\frac{\textbf{d}^{2}-3\frac{(\textbf{d}\textbf{r})^{2}}{r^{2}}}{r^{3}}+\frac{4\pi}{3}\textbf{d}^{2}\delta(\textbf{r})=-d^{\alpha}d^{\beta}\partial^{\alpha}\partial^{\beta}\frac{1}{r},\end{equation}
with $\partial^{\alpha}=\nabla^{\alpha}$ is the spatial derivative (the gradient operator). This the full potential of interaction of electric dipoles, the magnetic dipoles will be discussed below.
The delta-function term, which is the last term in the middle part of formula (\ref{FSDbec d-d Ham full}), appears due to next arguments.

Potential of the electric field caused by an electric dipole $\textbf{d}$ can be obtained as $\varphi=-(\textbf{d}\nabla)(1/r)$ (see for instance \cite{Landau 2}). Consequently the electric field appears as $\textbf{E}=-\nabla\varphi$$=(\textbf{d}\nabla)\nabla(1/r)$. In tensor notations it can be presented as $E^{\alpha}=d^{\beta}\nabla^{\beta}\nabla^{\alpha}(1/r)$. Energy of interaction of two electric dipoles appears $U_{dd}=-\textbf{d}_{2}\textbf{E}_{21}$, where $\textbf{d}_{2}$ is the electric dipole moment of the second dipole, and $\textbf{E}_{21}$ is the electric field caused by the first dipole acting on the second dipole. Putting explicit form of electric field created by the first dipole in formula for the potential energy we find $U_{dd}=-d^{\alpha}_{2}d^{\beta}_{1}\nabla^{\beta}\nabla^{\alpha}(1/r)$.
For understanding of this formula we should get explicit formula for the potential energy using well-known identity
\begin{equation}\label{NI GP togdestvo}-\partial^{\alpha}\partial^{\beta}\frac{1}{r}= \frac{\delta^{\alpha\beta}-3r^{\alpha}r^{\beta}/r^{2}}{r^{3}}+\frac{4\pi}{3}\delta^{\alpha\beta}\delta(\textbf{r}).\end{equation}
Therefore we have obtained formula (\ref{FSDbec d-d Ham full}).
The potential energy of dipole-dipole interaction can be represented in terms of the Green function $G^{\alpha\beta}$ of dipole-dipole interaction $U_{dd}=-d^{\alpha}_{2}d^{\beta}_{1}G^{\alpha\beta}$, where $G^{\alpha\beta}=\partial^{\alpha}\partial^{\beta}\frac{1}{r}$.
Full potential of dipole-dipole interaction is presented on Figs. (2) and (3). They appear as cross-section of Fig. (1) by plane perpendicular 0z axes (0x axes) for Fig. (2) (Fig. (3)), but including $\delta$ function in cross-sections passing though the point $x=z=0$. Figs. (2) and (3) also show finite radius of particles limiting area of potential existence. At increasing of $z$ on Fig. (2) ($x$ on Fig. (3)) value of $r_{0}$ should decrease due to spherical form of particles, but we have not shown it on figures keeping $r_{0}$ as a constant for illustration of contribution of the finite size of particles. The $\delta$ function term on Fig. (2) gives repulsion additional to very strong repulsion given by fraction (\ref{FSDbec d-d Ham fraction only}) and shown on Fig. (1). But the delta function term on Fig. (3) reveals repulsion at small distances for $\textbf{d}\parallel \textbf{r}$. This repulsion leads to stabilization of spectrum of collective excitations in electric dipolar BECs, as we also see from formulas below.

\begin{figure}
\includegraphics[width=8cm,angle=0]{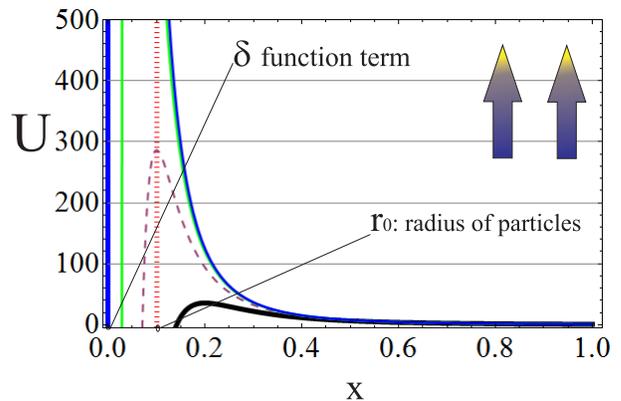}
\caption{\label{ESS 2} (Color online) The figure illustrates the dependence of dipole-dipole interaction potential
on $x$. Different lines are made for different $z$ at $y=0$: black (), dashed, green (), blue lines correspond decreasing of $z$. Lines at $z\neq 0$ are described by formula $U=d^{2}(x^{2}-2 z^{2})/(x^{2}+z^{2})^{5/2}$. The line at $z=0$, $y=0$ is described by next formula $U=d^{2}/\mid x\mid^{3}+4\pi d^{2}\delta(\textbf{r})/3$.}
\end{figure}
\begin{figure}
\includegraphics[width=8cm,angle=0]{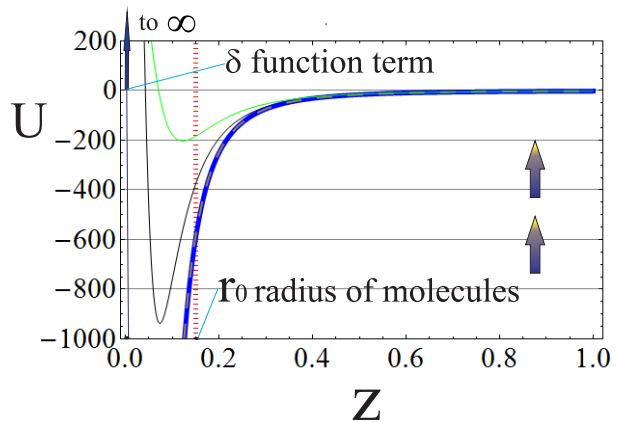}
\caption{\label{ESS 3} (Color online) The figure illustrates the
dependence of dipole-dipole interaction potential on $z$ at different $x$ and $y=0$: black, green, dashed, blue lines correspond decreasing of $x$ Dashed line almost coincides with the blue one, but the blue line has $\delta$ function addition.
Since the blue line at $x=0$, $y=0$ is described by next formula $U=-2d^{2}/\mid z\mid^{3}+4\pi d^{2}\delta(\textbf{r})/3$.
The blue vertical line at $z=0$ shows positive infinite potential giving repulsion.
It appears as continue of the attractive part given by fraction (\ref{FSDbec d-d Ham fraction only}).}
\end{figure}

Thus we have to include the delta-function term in the potential of dipole-dipole interaction. Let us admit that the semi-relativistic Darwin term describing two-particle interaction is proportional to $\delta$ function \cite{Landau 4} (see section 83 the second term in formula 83.15), \cite{Ivanov Darwin} (see formula 4). Nevertheless it gives considerable contribution in dispersion of the Langmuir waves in quantum plasmas \cite{Ivanov Darwin}, \cite{Asenjo NJP 12}. The semi-relativistic Darwin term is related to the Zitterbewegung \cite{Asenjo NJP 12}. It shows additional example of the delta-function term contributing in collective dynamics.

If we want to consider finite size of particles we lose the delta-function term,  but it is not enough to include the finite size. For account of finite size we should go further.
We need to introduce a finite radius $r_0$ of atoms and consider the integral describing dipole-dipole interaction not over all space, but over space outside of a sphere having radius $2r_0$.

The integral term usually existing in the gGP equation can be represented in non-integral form by explicit introduction of the electric field at account of the full potential of electric dipole interaction (see Refs. \cite{Andreev 2013 non-int GP}, \cite{Andreev 2013 Dip+Spin}). It is very useful for a point-like particle model
$$\imath\hbar\partial_{t}\Phi(\textbf{r},t)$$
\begin{equation}\label{FSDbec nlse polariz non Int ED} =\Biggl(-\frac{\hbar^{2}}{2m}\triangle+g\mid\Phi(\textbf{r},t)\mid^{2}-d\textbf{l}\textbf{E}(\textbf{r},t)\Biggr)\Phi(\textbf{r},t).\end{equation}
As the result the gGP equation becomes a non-integral equation containing new dynamical function: the electric field $\textbf{E}(\textbf{r},t)$.
An explicit form of the electric field is $E^{\alpha}(\textbf{r},t)=\int d\textbf{r}' G^{\alpha\beta}(\textbf{r},\textbf{r}')P^{\beta}(\textbf{r}',t)$, with $\textbf{P}(\textbf{r},t)=d\textbf{l} n(\textbf{r},t)$, where $\textbf{l}$ is a fixed direction of dipoles. This field satisfies the Maxwell equations
\begin{equation}\label{FSDbec div E}\nabla \textbf{E}(\textbf{r},t)=-4\pi d\textbf{l} \nabla n(\textbf{r},t),
\end{equation}
and
\begin{equation}\label{FSDbec curl E}\nabla\times \textbf{E}(\textbf{r},t)=0.
\end{equation}
Electric field caused by dipoles is introduced in formula (\ref{FSDbec div E}) for fully polarized dipolar BECs \cite{Andreev 2013 non-int GP}, \cite{Andreev 2013 Dip+Spin}. Electric field for more general case of dipolar BECs with evolution of directions of electric dipole moments was considered in Ref. \cite{Andreev EPJ D Pol}. It was also introduced in Ref. \cite{Wilson NJP 12}, but authors used common potential of the dipole-dipole interaction (\ref{FSDbec d-d Ham fraction only}). In formulas (\ref{FSDbec nlse polariz non Int ED})-(\ref{FSDbec curl E}) we have used the self-consistent field approximation for the long-range dipole-dipole interaction between point-like particles. There is a similar approach in plasma physics, where one use the self-consistent field approximation for quasi-static electric field created by electrons. This field obeys the Maxwell equations. This approach is called the Vlasov-Poisson approximation.

For consideration of particles with finite radius we need to get back to integral form of equations. At our choice we can consider the integral gGP equation or the corresponding hydrodynamic equations. We prefer to use hydrodynamic equations, which appears first at microscopical derivation and can exist even in cases when we can not obtain non-linear Schrodinger equation \cite{Andreev PRA08}, \cite{Andreev PRB 11}. They are to be
\begin{equation}\label{FSDbec continuity equation}\partial_{t}n+\nabla (n\textbf{v})=0,
\end{equation}
and
$$mn(\partial_{t}+\textbf{v}\nabla)\textbf{v}-\frac{\hbar^{2}}{2m}n\nabla\Biggl(\frac{\triangle n}{n}-\frac{(\nabla n)^{2}}{2n^{2}}\Biggr)$$
\begin{equation}\label{FSDbec Euler integral} =-gn\nabla n+d^{\beta}d^{\gamma}n\nabla\int
d\textbf{r}'G^{\beta\gamma}(\textbf{r},\textbf{r}') n(\textbf{r}',t).
\end{equation}
The continuity equation (\ref{FSDbec continuity equation}) contains the particles concentration $n$ and the velocity field $\textbf{v}$. The particle concentration has microscopic definition $n(\mathbf r, t)=\int d R \sum _{p=1}^N\delta(\mathbf r-\mathbf r_p)\Psi^\dag(R,t) \Psi(R,t)$, with $dR=\Pi_{p=1}^N d\mathbf r_p$ (see Refs. \cite{Andreev PRA08}, \cite{Andreev PRB 11}, they also contain derivations of the quantum hydrodynamic equations). The hydrodynamical variables are related with macroscopic wave function governed by the GP equation: $n=\mid\Phi\mid^{2}$, and $v=\hbar\mid\Phi\mid^{2}\nabla\phi/m$, where $\phi$ is the phase of the wave function $\Phi$, m is the mass of particles. The Euler equation consists of several terms. The first group of terms has kinematic nature. The second group of terms proportional to the square of the Plank constant $\hbar^{2}$ is the quantum Bohm potential, related to the de-Brougle nature of quantum particles. The right-hand side of equation (\ref{FSDbec Euler integral}) contains interparticle interaction presented by two terms related to two differen type of interactions. The short-range interaction is presented by the first term proportional $g$ appears in the first order by the interaction radius (the Gross-Pitaevskii approximation) \cite{Andreev PRA08}, where $g=4\pi\hbar^{2}a_{0}/m$ is the interaction constant. The last term in the right-hand side is the dipole-dipole interaction.
The integral in equation (\ref{FSDbec Euler integral}) is integral over whole space. Thus a point $\textbf{r}-\textbf{r}'$ is also included. It corresponds to point like particles. For consideration of finite radius of molecules we need to restrict area of integration taking integral over whole space except a sphere of radius $r_{0}=2r_{i}$ with center at $\textbf{r}-\textbf{r}'=0$, where $r_{i}$ is a radius of a molecule. If we have a system of parallel dipoles with fixed direction parallel $0z$ axes we can put $\alpha=\beta=z$ in the last term of equation (\ref{FSDbec Euler integral}).

\begin{figure}
\includegraphics[width=8cm,angle=0]{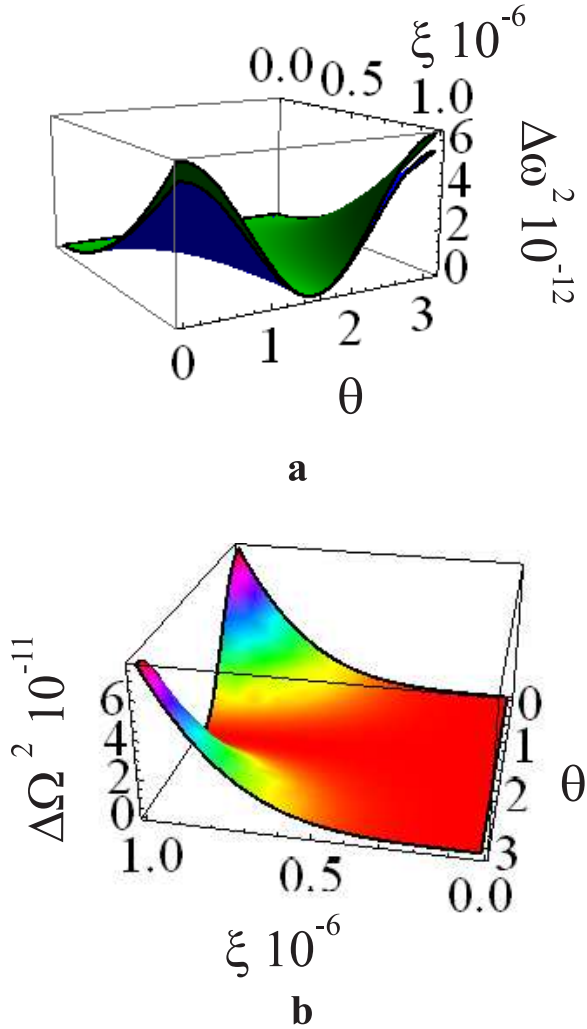}
\caption{\label{ESS 4} (Color online) The figure illustrates the
dipolar part of the dispersion dependence for the collective
excitations in the dipolar BECs. We picture shift of square of frequency in the Bogolubov's spectrum caused by dipoles
$\Delta\omega^{2}=\xi^{2}\cos^{2}\theta\cos(R\xi)$, with $\xi=k/\kappa$, and $R=2r_{0}\kappa$.
$\Delta\omega^{2}$ depicts the last term in formula (\ref{FSDbec disp for discuss}) describing dispersion of the electrically polarized BECs.
Figure (a) presents two surfaces giving the dispersion dependence on the angle $\theta$ between the direction of wave propagation and the direction of external electric field \emph{and} the reduced wave vector $\xi$. Upper of them (green) shows $\Delta\omega^{2}$ for zero radius particles. Lower surface gives dispersion dependence for finite size particles with radius $r_{0}=4$ $10^{-8}$  cm. We apply $n_{0}=10^{14}$ cm$^{-3}$, $m=2.11$ $10^{-22}$ g, $d=1$D$=3.3$ $10^{-30}$ C m $=10^{-18}$ CGS units giving $\kappa=0.4$ for both figures. Difference between two surfaces on figure (a) $\Delta\Omega^{2}=\Delta\omega^{2}(R=0)-\Delta\omega^{2}(R=2$ $10^{-7})$ is presented on figure (b) to show magnitude of contribution of finite radius effect on the dispersion dependence.}
\end{figure}
\begin{figure}
\includegraphics[width=8cm,angle=0]{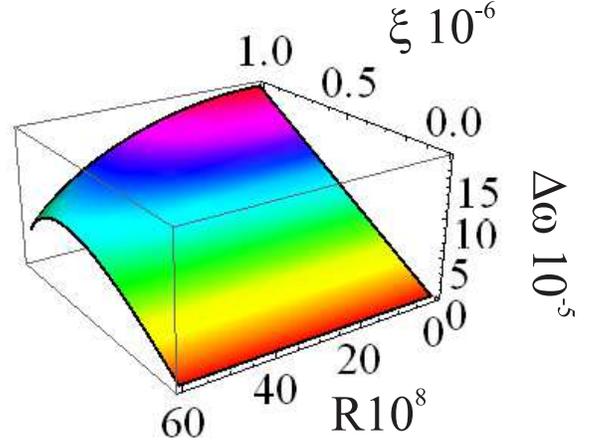}
\caption{\label{ESS 5} (Color online) This figure shows $\Delta\omega$ defined in notation to Figure (4) and presents dependence of $\Delta\omega$ on the reduced wave vector $\xi$ and the reduced radius of particles $R$ at $\theta=\pi/4$. We use here same parameters of the system as on Fig. (4).}
\end{figure}

Assuming particles have spherical shape and source of field distributed on surface we are getting spectrum of collective excitations. We consider that the equilibrium part of the particle concentration is a nonzero constant and the velocity field equals to zero in equilibrium. Considering small perturbations of the equilibrium $\delta n=N\exp(-i\omega t+i\textbf{k}\textbf{r})$ and $\delta \textbf{v}=\textbf{U}\exp(-i\omega t+i\textbf{k}\textbf{r})$ we can obtain spectrum $\omega(\textbf{k})$.

Our calculation gives the following spectrum of collective excitations
\begin{equation}\label{FSDbec disp for discuss}\omega^{2}=\frac{\hbar^{2}k^{4}}{4m^{2}}+\frac{gn_{0}k^{2}}{m}+\frac{4\pi n_{0}d^{2}k^{2}}{m}\cos^{2}\theta\cos\zeta,\end{equation}
where $\cos\theta=k_{z}/k$, and $\cos\zeta$ is the contribution of finite radius of molecules $r_{0}$ giving additional contribution in dispersion dependence $\zeta=2r_{0}k$. Getting to point-like particles we have $\zeta\rightarrow0$ and we come to the result we obtained earlier including delta-function term in potential of dipole-dipole interaction \cite{Andreev 2013 non-int GP}, \cite{Andreev 2013 Dip+Spin}. This limit differs from formula
\begin{equation}\label{FSDbec disp from Lit}\omega^{2}=k^{2}\Biggl(\frac{n_{0}}{m}\biggl(g+\frac{C_{dd}}{3}(3\cos^{2}\theta-1)\biggr)+\frac{\hbar^{2}k^{2}}{4m^{2}}\Biggr)\end{equation}
obtained in other papers (see, for example, formula (5.1) in Ref. \cite{Lahaye RPP 09} or formula (11) in Ref. \cite{Baranov CR 12}, $C_{dd}$ is the dipolar coupling constant $C_{dd}=4\pi d^{2}$), where authors have not considered the delta-function term in the dipole-dipole interactions following the model developed in basic papers \cite{Yi PRA 00}-\cite{Yi PRA 01}.

We can see from formula (\ref{FSDbec disp for discuss}) that in the limit of point-like particles $\zeta\rightarrow0$ formula reveals no roton instability for three dimensional electrically dipolar BECs. Stabilization of the spectrum in compare with formula (\ref{FSDbec disp from Lit}) appears due to account of the delta function in the potential of dipole-dipole interaction. The contribution of delta function cancels $-C_{dd}/3$ in formula (\ref{FSDbec disp from Lit}).

Main results of this letter is account of finite size of particles given by $\cos\zeta$ in formula (\ref{FSDbec disp for discuss}), where $\zeta$ is defined in terms of radius of particles $r_{0}$ and the module of wave vector $k$. Thus dependence on the finite size of molecules gives additional dependence on $k$. We do not include shape of molecules approximately considering them as spheres. $^{39}$K$^{85}$Rb is an example of ultracold Bose molecules have been used in experiment. More examples can be found in Ref. \cite{Quemener CR 12} (see Table 1 on page 4959).
We do not consider vicinity of point $\textbf{r}=0$. So we use the fact that tensor $(\delta^{\alpha\beta}-3r^{\alpha}r^{\beta}/r^{2})/r^{3}$ equals to $-\partial^{\alpha}\partial^{\beta}\frac{1}{r}$
in whole space except one point $\textbf{r}=0$. The Fourier image of
$-\partial^{\alpha}\partial^{\beta}\frac{1}{r}$ appears as $4\pi k^{\alpha}k^{\beta}/k^{2}$.

Figures (4.a) and (4.b) shows that finite radius change the dispersion dependence in area of large wave vectors $\xi=k/\kappa$, where $\frac{1}{\kappa^{2}}=\frac{4\pi d^{2}n_{0}}{m}$. It reveals at all angles except $\theta=\pi/2$, where the shift of $\omega^{2}$ in (\ref{FSDbec disp for discuss}) given by the polarization $\Delta\omega^{2}$ equals to zero for any $\xi$ and $R=2r_{0}\kappa$. Maximal difference in dispersion surfaces can be found at $\theta=0$ and $\pi$, where $\Delta\omega^{2}$ is maximal at any $R$. Fig. (4) shows that finite size of particle $r_{0}=4$ $10^{-8}$ cm gives contribution of order of 10 percents, decreasing $\omega(\textbf{k})$ at all angles. We also studied dependence of $\Delta\omega$ on radius of molecules at fixed angle $\theta=\pi/4$. It is shown on Fig. (5) presenting change of $\Delta\omega(\xi)$ with increasing of $R$. We have linear dependence for point-like particles. Giving finite size of particles we find it gets curvature to the dependence. Increasing $R$ gives decreasing of whole curve and reveals that it reach an extreme point of local maximum making $\Delta\omega(\xi)$ non-monotonic behaviour.

In classic physics the wave length of matter waves is limited by average interparticle distance. In quantum mechanics we hit the de-Broglie wave nature of particles, hence matter waves can continuously convert into collective quantum excitations with wave lengths smaller than interparticle distance.

New type of instability arise due to finite size of particles. It appears at large wave vector, which have not covered by Fig. (5). On Fig. (4) we have taken $\xi\leq 10^{6}$. Let us consider area of larger $\xi$ at $r_{0}=5$ $10^{-8}$cm. In this case $\Delta\omega(k)$ decreases at large $k$ and becomes negative at $k_{0}\cdot 2r_{0}\geq\pi/2$. Therefore at small positive interaction constant $g$ an instability can arise for $g+4\pi d^{2}\cos^{2}\theta\cos\zeta <0$. This short wavelength instability appears due to finite size of particles.

All these conclusions have been made for dipolar BECs with the electric polarization. We have shown \cite{Andreev 2013 Dip+Spin} that evolution of electrically and magnetically polarized BECs governs by different equations due to the fact that the delta function term in the potential of dipole-dipole interactions contains different coefficients in front of the delta function for electric and magnetic dipoles.

Magnetically dipolar BECs usually described by integral gGP equation and spinor BECs described by spinor GP equation containing Zeeman terms \cite{Ho PRL 98}-\cite{Stamper-Kurn RMP 13} are closely related topics describing same physical systems in different manner. It can be easily seen from next consideration. Non integral form of gGP equation (\ref{FSDbec nlse polariz non Int ED}) is written for electrically dipolar BECs. However similar equation can be written for magnetically dipolar BECs (see \cite{Andreev 2013 Dip+Spin} formula 13). In the magnetic case we have $-\mu\textbf{l}\textbf{B}(\textbf{r},t)$ instead of $-d\textbf{l}\textbf{E}(\textbf{r},t)$, where the magnetic field $\textbf{B}$ satisfies the following Maxwell equations $\nabla \textbf{B}=0$ and $\nabla\times \textbf{B}=4\pi\mu\nabla n\times \textbf{l}$, where $\mu$ is the magnetic moment of particles, and $\mu\nabla n\times \textbf{l}$ is the curl of magnetization $\textbf{M}$. Therefore we can see that the integral gGP equation appears in the form similar to the linear Zeeman term. But there is a difference as well. The Zeeman terms in the spinor BECs exist due to an external field and do not related to the magnetic field created by magnetic moments of the system. In our case the magnetic field is the sum of external and internal fields, where the last one appears from the integral term of the gGP equation. Let us notice that correct magnetic field appears together with the Maxwell equations at consideration of full magnetic dipole interaction with correct coefficient in front of the delta function $U_{\mu\mu}=-\mu^{\alpha}\mu^{\beta}(\partial^{\alpha}\partial^{\beta}\frac{1}{r}+4\pi\delta^{\alpha\beta}\delta(\textbf{r}))$ (see Ref. \cite{Andreev 2013 Dip+Spin} formula 15 and textbook \cite{Landau 4} section 83 the last group of terms in formula 15, where the Hamiltonian of spin-spin interaction appears as a part of the Breit Hamiltonian). However consideration of the spinor BECs includes evolution of magnetization direction as spinor Schrodinger (Pauli) equation is equivalent to three hydrodynamic equation: the continuity, Euler and magnetic moment evolution equation (see for instance \cite{Andreev Asenjo 13}). It has not included in formula (13) of Ref. \cite{Andreev 2013 Dip+Spin}. Let us admit that the $\delta$ function term for electric dipolar BECs gives an isotropic repulsion, when the $\delta$ function term for magnetic dipolar BECs give an isotropic attraction.

Let us briefly describe consequences of finite size of particles for magnetic dipolar BECs. The spectrum of collective excitation in fully polarized magnetic dipolar BECs for point-like particles with the full potential of dipole-dipole interaction was recently obtained in Ref. \cite{Andreev 2013 Dip+Spin} (see formula (28)). Here we show how it changes due to finite size of atoms. Neglecting the delta function term in the potential of spin-spin interaction we get formula (\ref{FSDbec d-d Ham fraction only}). Considering whole space except vicinity of point $\textbf{r}=0$ we get spectrum (\ref{FSDbec disp for discuss}) for magnetized BECs as well. It dramatically changes properties of the spectrum. Polarization term in magnetically dipolar BECs is negative for all angles (see Ref. \cite{Andreev 2013 Dip+Spin} formula (28)) unlike electric dipolar BECs. This difference appears due to different coefficients in front of the delta-function in the potential of dipole-dipole interaction. So magnetically dipolar BECs reveals the roton instability (for small positive $g$) along with the phonon instability (for negative $g$). Neglecting the delta function term changes sign of magnetic dipole contribution making it similar to finite size spectrum of electric dipoles (\ref{FSDbec disp for discuss}).

As a conclusion we have pointed out that developed equations open possibilities for getting contribution of size of atoms and molecules on various properties of dipolar BECs. We have used this it for studying of fundamental spectrum of linear excitations for illustration of main consequences.


\begin{thebibliography}{17}

\bibitem{Lahaye Nature 07} T. Lahaye et. al., Nature \textbf{448}, 672 (2007).

\bibitem{Giamarchi NP 08} T. Giamarchi, C. Ruegg
and O. Tchernyshyov, Nature Phys. \textbf{4}, 198 (2008).

\bibitem{Ni PCCP 09} K.-K. Ni, S. Ospelkaus, D. J. Nesbitt, J. Ye and D. S. Jin, Phys. Chem. Chem. Phys. \textbf{11}, 9626 (2009).

\bibitem{Carr NJP 09} L. D. Carr et al., New J. Phys. \textbf{11}, 055049 (2009).


\bibitem{Yi PRA 00} S. Yi and L. You, Phys. Rev. A, \textbf{61}, 041604(R) (2000).

\bibitem{Goral PRA 00} K. Goral, K. Rzazewski, and T.
Pfau, Phys. Rev. A \textbf{61}, 051601(R) (2000).

\bibitem{Santos PRL 00} L. Santos, G.V. Shlyapnikov, P. Zoller, and M.
Lewenstein, Phys. Rev. Lett. \textbf{85}, 1791 (2000).

\bibitem{Yi PRA 01} S. Yi and L. You, Phys. Rev. A, \textbf{63}, 053607 (2001).


\bibitem{Jona-Lasinio 13} M. Jona-Lasinio, K.  Lakomy and L. Santos, Phys. Rev. A \textbf{88}, 013619 (2013).

\bibitem{Adhikari JP B 14} S. K. Adhikari and Luis E. Young-S., J. Phys. B \textbf{47},
015302 (2014).

\bibitem{Koberle PRA 12} P. Koberle, D. Zajec, G. Wunner, and B. A. Malomed, Phys. Rev. A \textbf{85}, 023630 (2012). 


\bibitem{Szirmai PRA 12} G. Szirmai and P. Szepfalusy,
Phys. Rev. A \textbf{85}, 053603 (2012).

\bibitem{Szirmai EPJ ST 13} G. Szirmai, The Eur. Phys. J. Special Topics
\textbf{217},  189 (2013).

\bibitem{Ticknor PRL 11} C. Ticknor, R.M. Wilson, and J.L. Bohn, Phys. Rev. Lett. \textbf{106}, 065301 (2011).

\bibitem{Wilson PRA 12} R. M. Wilson, C. Ticknor, J. L. Bohn, and E. Timmermans,
Phys. Rev. A \textbf{86}, 033606 (2012).



\bibitem{Andreev 2013 non-int GP} P. A. Andreev, Mod. Phys. Lett. B \textbf{27}, 1350096 (2013).

\bibitem{Andreev EPJ D Pol} P. A. Andreev and L. S. Kuz'menkov, Eur. Phys. J. D \textbf{67}, 216 (2013).

\bibitem{Andreev 2013 Dip+Spin} P. A. Andreev, L. S. Kuz'menkov, arXiv:1303.7362.


\bibitem{Landau 2} L. D. Landau and E. M. Lifshitz, \textit{The Classical Theory of Fields} (Butterworth-Heinemann, 1975).

\bibitem{Landau 4} V. B. Berestetskii, E. M. Lifshitz, L. P. Pitaevskii,
Vol. 4 \textit{Quantum Electrodynamics} 
(Butterworth-Heinemann, 1982).

\bibitem{Ivanov Darwin} A. Yu. Ivanov, P. A. Andreev, L. S. Kuz'menkov, arXiv: 1209.6124.

\bibitem{Asenjo NJP 12} F. A. Asenjo, J. Zamanian, M. Marklund, G. Brodin,
and P. Johansson, New J. Phys. \textbf{14}, 073042 (2012).


\bibitem{Wilson NJP 12} R. M. Wilson, S. T. Rittenhouse and J. L. Bohn, New J. Phys. \textbf{14}, 043081 (2012).

\bibitem{Andreev PRA08} P. A. Andreev, L. S. Kuz'menkov, Phys. Rev. A \textbf{78}, 053624 (2008).

\bibitem{Andreev PRB 11} P. A. Andreev, L. S. Kuzmenkov, M. I. Trukhanova, Phys. Rev. B \textbf{84}, 245401 (2011).



\bibitem{Lahaye RPP 09} T. Lahaye, C. Menotti, L. Santos, M. Lewenstein, and T. Pfau, Rep. Prog. Phys. \textbf{72}, 126401 (2009).
\bibitem{Baranov CR 12} M. A. Baranov, M. Dalmonte, G. Pupillo, and P. Zoller, Chem. Rev.,  \textbf{112},  5012 (2012).




\bibitem{Quemener CR 12} Goulven Quemener, and Paul S. Julienne, Chem. Rev. \textbf{112}, 4949 (2012).

\bibitem{Ho PRL 98} T. L. Ho, Phys. Rev. Lett. \textbf{81}, 742 (1998).
\bibitem{Machida JPSJ 98} T. Ohmi and K. Machida, J. Phys. Soc. Jpn. \textbf{67},
1822 (1998).

\bibitem{Kawaguchi Ph Rep 12} Yuki Kawaguchi, Masahito Ueda, Physics Reports \textbf{520},  253 (2012).
\bibitem{Stamper-Kurn RMP 13} Dan M. Stamper-Kurn, Masahito Ueda, Rev. Mod. Phys. \textbf{85}, 1191 (2013).

\bibitem{Andreev Asenjo 13} P. A. Andreev,
F. A. Asenjo, and S. M. Mahajan, arXiv: 1304.5780.

%
\end{thebibliography}
\end{document}